\documentstyle[epsfig,amssymb,floatflt,float,here]{mn2e}

\newif\ifAMStwofonts

\newcommand{\sm}{\, {\rm M}_{\odot}}

\newcommand{\kms}{km~s$^{-1}$}

\newcommand{\ndeg}{^{\circ}}

\title[Is the dark halo of our Galaxy spherical?]{Is the dark halo of
our Galaxy spherical?} \author[A. Helmi] {Amina Helmi
\\ Astronomical Institute Utrecht, P.O.Box 80000,
3508 TA Utrecht, and \\ Kapteyn Institute, P.O.Box 800, 9700 AV
Groningen, The Netherlands. E-mail:ahelmi@astro.rug.nl}

\pagerange{\pageref{firstpage}--\pageref{lastpage}}
\pubyear{}

\begin{document}

\maketitle

\label{firstpage}

\begin{abstract}
It has been recently claimed that the confined structure of the debris
from the Sagittarius dwarf implies that the dark-matter halo of our
Galaxy should be nearly spherical, in strong contrast with predictions
from cold dark-matter (CDM) simulations, where dark halos are found to
have typical density axis ratios of $0.6 - 0.8$. In this letter,
numerical simulations are used to show that the Sagittarius streams
discovered thus far are too young dynamically to be sensitive to the
shape of the dark halo of the Milky Way. The data presently available
are entirely consistent with a Galactic dark-matter halo that could
either be oblate or prolate, with minor-to-major density axis ratios
$c/a$ as low as 0.6 within the region probed by the orbit of the
Sagittarius dwarf.
\end{abstract}


\section{Introduction}

The relevance of measuring the shapes of dark-matter halos lies in the
fact that they are sensitive to the nature of dark-matter particles
themselves. Depending on whether these are cold, self-interacting or
hot, the shape of a dark halo may be flattened (e.g. Dubinski \&
Carlberg 1991), close to spherical (Yoshida et al.\ 2000; Dav\'e et
al.\ 2001) or spherical (Mayer et al. 2002). If dark-matter was
baryonic, then the halos are expected to be strongly flattened, with a
ratio of minor to major axis of about 0.2 (Pfenninger \& Combes
1994). If, on the other hand, the observed flat rotation curves are
due to a modified law of gravity, i.e. MOND, then the ``fictitious
dark component'' ought to be equivalent to a spherical (massless) halo
(Milgrom 2001).

Predictions for the shapes of dark-matter halos come mostly from
N-body simulations, although some analytic estimates using ellipsoidal
collapse models are also available (Eisenstein \& Loeb,
1995). Numerical studies using the Standard CDM power spectrum have
often yielded triaxial halos with minor-to-major axis ratios of $c/a
\sim 0.4 - 0.6 $, with cluster size halos more flattened than galaxy
size halos (Frenk et al. 1988; Warren et al. 1992; Thomas et
al. 1998). More recent studies in a $\Lambda$CDM cosmology, have shown
the typical axis ratios to be $\sim 0.6 - 0.8$ for Milky Way size
halos (Bullock 2002).

Observationally, the measurement of the shape of a dark-matter halo is
a non-trivial task, and as such different techniques using a variety
of tracers have found notably disparate values (see Sackett 1999 for a
good review). Powerful probes of the shape of dark halos are tidal
streams, since they consist of stars moving on very similar orbits
(Lynden-Bell \& Lynden-Bell 1995; Johnston 1998). For example, the
plane of motion of a tidal stream moving in a spherical potential
remains constant in time, implying that its stars will be located on a
great circle arc on the sky.  Also the width of such a stream will
remain constant in an average sense; it will, however, vary
periodically depending on the location of its stars along their orbit
(Helmi et al. in preparation).

In the case of the Milky Way, Ibata et al. (2001) have concluded that
the recently discovered streams from the Sagittarius dwarf galaxy,
favor a spherical dark halo, $c/a \sim 1$, with oblate halos with
flattening values of $c/a \le 0.7$ rejected at very high confidence
levels (less than 1 chance in one million). Mayer et al. (2002) have
also claimed that the mere existence of tidal features around the
Milky Way and other nearby galaxies rules out strongly aespherical
($c/a \le 0.8$) and triaxial halos, since such features would disperse
very rapidly in flattened systems. More recently, Majewski et
al. (2003), using M giants from the 2MASS data were able to trace the
debris from the Sgr dwarf along an almost complete great circle arc on
the sky. The narrowness of the stream and its confinement also led
them to suggest a nearly perfectly spherical halo for the Milky Way.

One should, however, bare in mind that most of the Sgr debris
discovered thus far is probably constituted by material lost from the
dwarf galaxy one to three passages ago, i.e. 1 to 3 Gyr ago, as
suggested by the models by Helmi \& White (2001) and Martinez-Delgado
et al. (2004). As will be shown below such dynamically young debris
streams may not be as sensitive to the shape of our Galaxy's dark halo
as previously thought.

In this letter, numerical simulations are used to model the evolution
of the Sgr dwarf galaxy and of its debris as it orbits the Milky
Way. The goal is to determine whether the presently available data on
the debris from Sgr can constrain the shape of the Galactic halo as
strongly as has been suggested in previous work. With this in mind, we
perform several experiments in which the Galactic potential includes a
dark halo component with different degrees of flattening. These
simulations are introduced in the Section 2. The analysis and results
of these simulations are presented in Sec.~3, and we conclude with a
brief discussion section.

\section{The simulations}

\begin{figure}
\includegraphics[height=8cm]{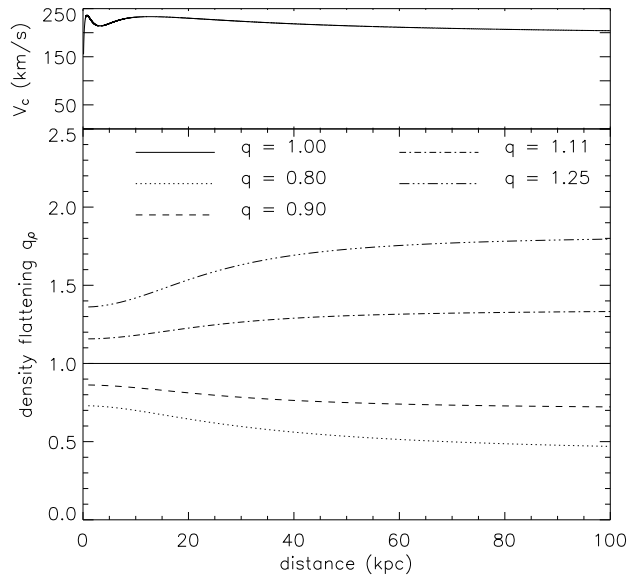}
\caption{Circular velocity along the major axis of our Galaxy model
(top panel). Variation of the dark halo flattening, measured by the
density axis ratio, as a function of distance from the centre of our
Galaxy model (bottom panel).}
\label{fig:q_rho}
\end{figure}

In our numerical simulations, we represent the Galaxy by a fixed potential
with three components: a dark logarithmic halo 
\begin{equation}
\label{sag_eq:halo}
\Phi_{\rm halo} = v^2_{\rm halo} \ln (R^2 + z^2/q^2 + d^2),
\end{equation}
a Miyamoto-Nagai disk
\begin{equation}
\label{sag_eq:disk}
\Phi_{\rm disk} = - \frac{G M_{\rm disk}}{\sqrt{R^2 + (a_{\rm d} + 
\sqrt{z^2 + b_{\rm d}^2})^2}},
\end{equation}
and a spherical Hernquist bulge
\begin{equation}
\label{sag_eq:bulge}
\Phi_{\rm bulge} = - \frac{G M_{\rm bulge}}{r + c_{\rm b}}, 
\end{equation}
where $d =12$ kpc and $v_{\rm halo}= 131.5$ \kms ; $M_{\rm disk} =
10^{11} \sm $, $a_{\rm d}$ = 6.5 kpc and $b_{\rm d}$~=~0.26 kpc;
$M_{\rm bulge} = 3.4 \times 10^{10} \sm$ and $c_{\rm b}$~=~0.7
kpc. This choice of parameters gives a flat rotation curve with an
asymptotic circular velocity of 186 \kms along the mayor axis of the
galaxy, and a circular velocity at the location of the Sun of 229
\kms, as shown in the top panel of Fig.\ref{fig:q_rho}. The parameter
$q$ is allowed to vary from 0.8 to 1.25, that is, from an oblate to a
prolate configuration. In the bottom panel of Figure \ref{fig:q_rho}
we show the variation of the density flattening for the halo component
in these models. Here $q_\rho$ is defined by the ratio of the
distances down the $z$ and $R$ axes at which a given isodensity
surface cuts those axes (see Binney \& Tremaine, 1987, p.~48).

The orbit of Sagittarius is relatively well constrained (Ibata et
al. 1997).  The heliocentric distance $d \sim 25 \pm 2$ kpc, position
$(l,b) = (5.6\ndeg, -14\ndeg)$, and the heliocentric radial velocity
$v_r^{\rm sun} \sim 140 \pm 2$ \kms of the galaxy core are well
determined. Recent proper motion measurements imply $v_\beta \sim 280
\pm 20$ \kms (Ibata et al. 2001). The strong North-South elongation of
the system, as well as its debris, suggest that the orbit should be
close to polar. To set up the simulations we generate random orbits
consistent with these constraints. We also require that the
present-day inclination of the orbit defined as $L_z/L$ (the ratio of
the $z$-component to the total angular momentum) does not deviate by
more than 5\% from that of a plane defined by the cross product of the
current position of Sgr ${\bold r}_{\rm Sgr} $ and the northern
${\bold r}_{\rm NS}$ (and southern ${\bold r}_{\rm SS}$) clumps of
debris detected by the SDSS (Ivezic et al. 2000; Yanny et al. 2000)

We represent the Sgr galaxy by a collection of $5 \times 10^4$
particles and model their self-gravity by a multipole expansion of the
internal potential to fourth order (White 1983; Zaritsky \& White
1988; see also Helmi \& White 2001). For the stellar distribution of
the pre-disruption dwarf a King model is chosen, since this has been
shown to fit well the dwarf spheroidals satellites of the Milky Way
(Mateo 1998). This model is specified by a combination of three
parameters: the depth of the potential well of the system
$\Psi(r\!=\!0)$, a measure of the central velocity dispersion
$\sigma^2$, and the central density $\rho_0$ or the King radius $r_0$.

In summary, we proceed as follows. For the spherical halo case, we
select an orbit as described above, integrate this orbit backward in
time (until $t_i = - 10$ Gyr), to provide the initial position and
velocity of the centre of mass of the satellite in the numerical
simulation. We then probe the parameter space of initial models (we
leave $\Psi(r\!=\!0)/\sigma^2 = 3.3 $ and $r_0 = 0.5$ kpc fixed),
until we find one that reproduces currently available observations of
the main body and of the debris from Sgr. It turns out that this
initial model is close to the purely stellar model of Helmi \& White
(2001). For the oblate and prolate halo potentials, we select orbits
which have similar mean apocentric and pericentric distances, as well
as similar $L_z$, to be able to determine what the effect of varying
$q$ is in a direct manner. The initial models for the satellite are
chosen so that, after 10 Gyr of evolution, the system has reached a
similar degree of disruption as in the spherical halo case, which is
typically set to 10 - 20 \%. The orbital properties and models of the
satellite are listed in Table~\ref{tab:ic}.

The set of stringent conditions imposed on the orbits and initial
models of the satellite are required for a fair comparison of the
effect of oblate and prolate halos on the properties of its debris.
Note that the aim is not so much to find the best fit model of the Sgr
dwarf and of its orbit for each one of the given potentials of the
Milky Way, but to compare similar orbits and models to get a handle on
how the variation in the degree of flattening affects their
evolution. The models presented here do reproduce the properties of
the main body of Sgr, as well as the location of its debris. This
implies that variations in the characteristics of the debris streams
may only be attributed to the variation in the shape of the dark halo
component.

\begin{table}
\begin{tabular}{ccccccc}
\hline
$q$ & ${\langle q_\rho \rangle} $ & $\langle R_{apo} \rangle $ & $\langle R_{peri} \rangle$ & 
$L_z$ & $\sigma$  & $M$ ($10^{8} M_\odot$)\\
\hline
0.80 & 0.60 & 59.2 & 10.1 & -1105.6 & 38 &  6.89 \\
0.90 & 0.78 & 56.4 & 13.7  & -911.9  & 29 &  4.01\\
1.00 & 1.00 & 53.9 & 14.5 & -938.5 & 28 &  3.74\\
1.11 & 1.26 & 51.4 & 14.0 & -929.0 & 27.5 &  3.61\\
1.25 & 1.63 & 52.3 & 14.7 & -967   & 25 &  2.98\\
\hline
\end{tabular}
\caption{Characteristic parameters of the orbits and initial models of
Sgr for the different degrees of flattening of the Galactic halo
assumed in our simulations. The second column is the mean flattening
of the dark halo component within the region probed by the orbit of
Sgr. Distances are in kpc, and velocities in \kms. In all models we
have fixed $\Psi(r\!=\!0)/\sigma^2 = 3.3 $ and $r_0 = 0.5$ kpc.}
\label{tab:ic}
\end{table}

\begin{figure}
\includegraphics[width=8.5cm,clip]{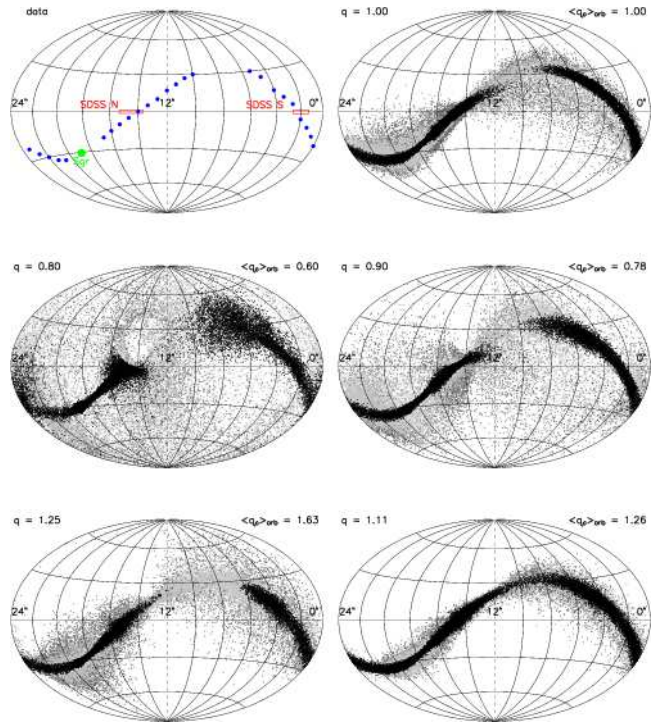}
\caption{The different panels show the sky distribution ($\alpha$,
$\delta$) of particles in our numerical simulations with varying
degrees of flattening for the dark halo of the Galaxy. The different
colours represent sets of particles which have become unbound more
than 6.5 Gyr ago (dark grey), more than 3 Gyr but less than 6.5 Gyr
ago (light grey), and less than 3 Gyr ago (black). In the top left
panel, the different detections of debris from Sgr are shown. Here the
black solid dots represent approximately the 2MASS M giants stream
(Majewski et al. 2003), while the boxes correspond to the northern and
southern detections by SDSS (Ivezic et al. 2000; Yanny et al. 2000),
which overlap substantially with the detections of the Spaghetti
Project Survey (Dohm-Palmer et al. 2001), QUEST (Vivas et al. 2001),
Martinez-Delgado et al. (2001), and Dinescu et al. (2002), which are
not plotted here for clarity.  Direct comparison of the first panel
with any of our numerical experiments suggests that the observations
could be tracing {\em only} the most recently formed streams of Sgr
(black dots in the different panels). See Figure~\ref{fig:dist} for
direct evidence that this is indeed the case. Such streams are too
young dynamically to have experienced the effects of a flattened halo,
be it prolate or oblate, and therefore remain quite coherent even when
the minor-to-major density axis ratios $c/a$ are as low as 0.6 within
the region probed by the orbit of Sgr, like in the 3rd., $\langle
q_\rho \rangle = 0.6$, and 5th., $\langle q_\rho \rangle = 1.63$,
panels of this Figure.}
\label{fig:sky}
\end{figure}

\section{Results}

The panels in Figure~\ref{fig:sky} show the sky distribution in
equatorial coordinates ($\alpha$, $\delta$) for all the particles in
the different experiments. The particles are colour-coded according to
when they became unbound from their parent satellite. Dark grey
corresponds to particles unbound in the first 3.5 Gyr of evolution
(approximately 5 pericentric passages), light grey to those released
between 3.5 and 7 Gyr (that is, between approximately 5 and 9
passages), while black are those that have been released in the last 3
Gyr of evolution (or less than 4 passages ago). This last subset also
includes particles that are still bound, i.e. what we call the main
body of Sgr.

It immediately strikes the eye that the distribution of black
particles is very similar in all five cases. The particles define
relatively narrow structures on the sky. These streams are so young
dynamically that they have not really been able to experience the
flattening of the dark halo in which they orbit, which eventually will 
lead to their thickening. Therefore, such young streams are not
sensitive enough to measure nor constrain the shape of a halo.

The precession and thickening effects, are on the other hand, visible
for the oldest streams in the most oblate halo, with $q = 0.8$ and an
average density flattening within the region probed by Sgr orbit of
$\langle q_\rho \rangle = 0.6$. Interestingly, in the case of the
prolate halo of similar flattening ($1/q=0.8$ and $\langle q_\rho
\rangle = 1.63$), the oldest streams are much more coherent. This is
probably due to the fact that the orbit of Sgr has a high-inclination,
i.e. it is essentially aligned with the principal axis of the density
distribution, and hence is less subject to strong precession.

For comparison, the first panel of Figure~\ref{fig:sky} shows the
distribution of claimed detections of debris from Sgr discovered so
far by the SDSS (Ivezic et al. 2000; Yanny et al. 2000; Newberg et
al. 2002) and by Majewski et al. (2003). There seems to be a very good
correspondence between the location of this debris and that of the
most recently lost particles in our simulations (black dots in the
remaining panels). Therefore, our results would seem to imply that the
observed streams could be, in fact, made up of stars lost by the Sgr
dwarf in the last 3 Gyr of evolution. If this were the case, the
immediate conclusion would be that such observations cannot put
meaningful constraints on the shape of the Galactic halo.

\begin{figure}
\includegraphics[width=8.5cm,clip]{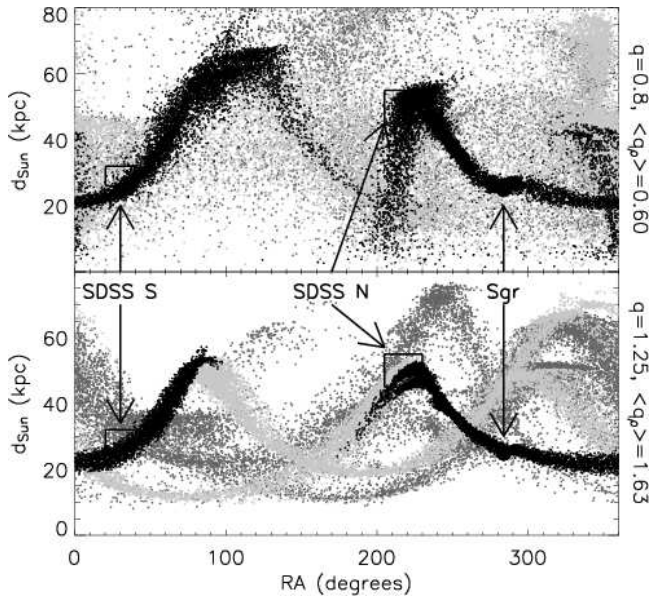}
\caption{The right ascension vs distance from the Sun distribution of
particles in our numerical simulations with $q=0.8$ (most oblate; top
panel) and $1/q = 0.8$ (most prolate; bottom panel). The same colour
coding as in Fig.\ref{fig:sky} is used. Also shown are the northern
and southern detections of debris from Sgr by SDSS. Note that these
overlap spatially (in the sky and in distance, as well as in radial
velocity) with the most recently unbound material, indicated here by
the black solid dots. This confirms that the streams discovered thus
far are dynamically young.}
\label{fig:dist}
\end{figure}

But are the black dots in Figure~\ref{fig:sky} really tracing {\em
all} the debris that has been discovered thus far? To determine
whether this is the case, we also need to check that their dynamical
properties (radial velocities and spatial location) are indeed
similar. To that end, in Figure~\ref{fig:dist} we show the distance
vs. right ascension distribution of particles for two of our
experiments: the most oblate and the most prolate cases. We have used
the same colour coding as before, and also overplotted the detections
of the northern and southern Sgr debris discovered by SDSS. It becomes
now clear that these detections are {\em all} tracing {\em only} the
material that has been released by the Sgr dwarf in the last 3
Gyr. All our experiments of different flattening essentially show the
same features, and are in agreement with the observational data
available about Sgr and its streams.

We also note here that the tidal streams traced by the 2MASS M giants
in Majewski et al.\ (2003), are also well-represented by the black
dots shown in Fig.~\ref{fig:sky} and Fig.~\ref{fig:dist}. 

The light grey dots representing particles released between 3.5 and 7
Gyr ago, are distributed along a thicker great circle path on the sky
in all but the most oblate halo experiments. Their sky location
overlaps substantially with the dynamically young streams and with the
observations of the streams from Sgr detected thus far. However, when
the additional dynamical information that is available is used, it
becomes clear that the overlap is only due to a projection effect, and
that these intermediate age streams have a different 3-dimensional
spatial distribution, as evidenced by their different distance
distribution as shown in Fig.~\ref{fig:dist}. There are currently no
observational data that would correspond to these intermediate age
streams, although see Dohm-Palmer et al. (2001) for a possible hint.

\section{Discussion}

We have carried out numerical experiments of the evolution of the Sgr
dwarf and its debris orbiting in Galactic potentials with dark halo
components with different degrees of flattening. These experiments
show that dynamically young streams (younger than about 3 Gyr) on a
Sgr-like orbit remain very coherent spatially, defining great circle
arcs on the sky even when the mean minor-to-major density axis ratio
$c/a$ of the dark halo of our Galaxy model within the region probed by
Sagittarius orbit is as low as 0.6 (prolate and oblate cases). A
straightforward comparison to the various detections of debris
material from Sgr leads to the unavoidable conclusion that this debris
is dynamically young. Hence its striking confinement to a great circle
arc on the sky does not imply that the dark halo of our Galaxy is
nearly spherical. This is in agreement with the conclusions of
Martinez-Delgado et al. (2004).

These results would seem to be in contradiction with those of Ibata et
al. (2001). The difference probably lies in the fact that, in their
simulations, the dwarf is completely disrupted by the present time (no
core can be seen in any of the figures corresponding to flattened mass
distributions). Hence they are unable to distinguish streams as
dynamically young as ours. This can either be due to the fact that the
chosen orbits were not a good representation of the actual orbit of
Sgr, or to their initial model of the dwarf, upon which they probably
did not impose any restrictions concerning its survival. 

It is perhaps worth emphasizing that to produce similar end products
(i.e. main body of the Sgr dwarf) in simulations with different
potential shapes, adjustments in the initial model of the dwarf are
necessary. For comparison, Ibata et al. (2001) use only two models in
their experiments. These are the models that evolved in a spherical
potential reproduce the properties of the main body (these were
presented in the papers by Ibata \& Lewis 1998; and Helmi \& White
2001).  When these initial models of the Sgr dwarf were evolved in
potentials with different degrees of flattening, they gave rise to
quite different present-day dwarfs: the evolution of the dwarf galaxy
is not independent of the potential in which it orbits. In contrast to
the approach adopted here, Ibata et al. did not ``correct" their
models for this effect and this is why their dwarf in the flattened
potentials does not survive until the present time, nor produces
dynamically young streams that would be visible today.

\begin{figure}
\includegraphics[width=8.2cm]{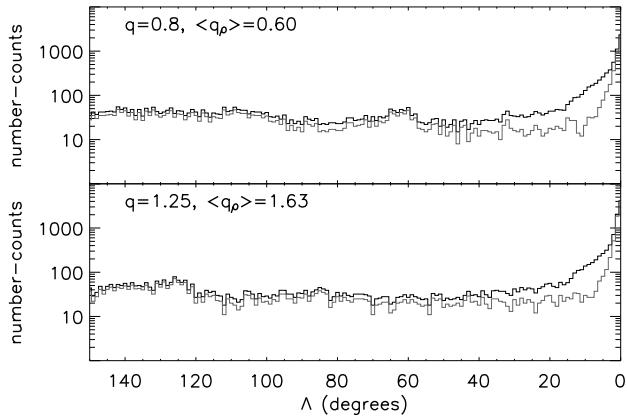}
\caption{Number counts of particles located within 5 kpc of the plane
defined by the Sgr dynamically young debris, as a function of angular
distance along the trailing stream $\Lambda$ (see for more details
Majewski et al. (2003)). The grey histogram corresponds to the actual
number counts, while the black curve includes the effect of a
background as measured by Majewski et al. (2003). There is good
agreement between the black histograms (both for $q=0.8$ and $q =
1.25$) and the data shown in Fig.13 of Majewski et
al. (2003). However, the model dwarf (background subtracted) appears
to be more concentrated than Majewski et al. (2003) suggest, although
its characteristics are consistent with previous observations of
Sgr. }
\label{fig:counts}
\end{figure}

It becomes clear that the initial conditions imposed on the orbit as
well as on the model of the progenitor of Sagittarius are both
critical to the interpretation of the numerical experiments run under
different dark matter halo shapes. Is it possible to constrain them in
a realistic way? Concerning the initial models of the dwarf,
observations of the main body of Sgr can be used as boundary
conditions. It is desirable that the observed characteristics of Sgr
are reproduced before one addresses the properties of its debris. All
models discussed in this paper satisfy this requirement in the
following sense. They reproduce the properties of the main body
described by Ibata et al.(1997), in terms of the extent, surface
brightness and internal velocity dispersion. A good correspondence is
also found for the number counts and the contrast between the main
body and the M giant streams. This is shown in Fig.\ref{fig:counts},
and should be compared to Fig.13 of Majewski et al. (2003).

The main difference between the stellar model discussed in Helmi \&
White (2001) and the models presented here, lies in the orbital
parameters (slightly larger apocentre and pericentre distances; these
were less well constrained three years ago). The mass ratio of the
remnant to the initial dwarf ranges between 10-20\% in all cases. The
fraction of mass contained in the dynamically youngest streams in the
stellar model of Helmi \& White (2001) is $\sim$ 10\%, while for the
models presented in this paper it varies from 13\% to 25\% of the
initial mass of the system. This implies slightly different recent
mass-loss rates for each one of the models presented here.  However,
the similarity in the spatial distribution of the youngest streams in
all our experiments indicates that this is not strongly dependent on
the detailed recent mass-loss history. The spatial distribution of
stars in these dynamically young streams is mostly determined by the
recent orbital trajectory, which by construction is similar in all
cases (see also Martinez-Delgado et al. 2004).

\begin{figure*}
\includegraphics[height=17cm,angle=270]{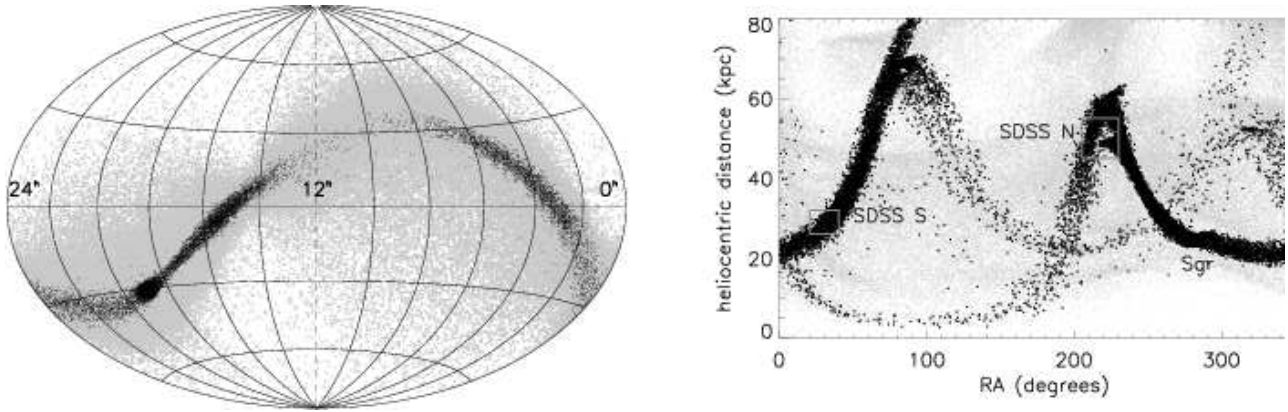}
\caption{The sky and right ascension vs distance from the Sun
distribution of particles in the ``stars + dark-matter'' model of
Helmi \& White (2001). Stars are shown in black, while dark-matter
particles are grey. Even in a dark-matter dominated system like that
shown here, there is a good correspondence between the properties of
the streams observed and those that are dynamically the youngest.}
\label{fig:DM}
\end{figure*}

We have chosen to model the dwarf using a King profile with only one
component of stellar nature. We have not taken into account the effect
of a dark-matter halo encompassing the stellar distribution, which
would have been important during the initial stages of evolution of
the dwarf (see Jiang \& Binney 2000). However, it is very likely that
the main body of the present day Sagittarius represents only the
central region of the progenitor dwarf (Helmi \& White 2001); and we
may assume that the dominant component of this central region is
stellar. This implies that the properties of the streams formed in the
last pericentric passages, i.e. those that are dynamically young, will
not to differ substantially from those originated in dark-matter
dominated progenitors. To support this statement, in Fig.\ref{fig:DM}
we have plotted the distribution of ``star-like'' particles for the
dark-matter dominated model of Helmi \& White (2001). The stellar
component of this model initially is very similar to the purely
stellar model discussed in the same paper (see their Fig. 6). The
mass-to-light ratio varies with radius (more precisely with binding
energy), implying that the streams formed at early times are
dark-matter dominated, while those dynamically young are purely
stellar.

The left panel of Fig.\ref{fig:DM} shows the sky distribution of
``dark-matter'' (in grey) and of ``star'' particles (in black). The
panel on the right shows their right ascension vs distance
distribution (same colour coding). These plots should be compared to
Figs.\ref{fig:sky} and \ref{fig:dist}. We note that the dynamically
young streams produced in a dark-matter dominated system, whose
initial stellar distribution is similar to that of the purely stellar
systems discussed in this paper, are not significantly different to
those produced in purely stellar systems. This is the case because the
streams highlighted here, and which we believe correspond to the
streams observed, are formed by material lost in the last few
pericentric passages and originate in regions of the initial system
which were dominated by stars (rather than by dark-matter). The
situation would perhaps be different if the Sgr dwarf was a system in
dynamical equilibrium, currently surrounded by a massive dark
halo. The most recently formed streams, would then have a large-dark
matter content. If however, as most authors conclude (Johnston,
Spergel \& Hernquist 1995; Velazquez \& White 1995; Zhao 1998;
G\'omez-Flechoso, Fux \& Martinet 1999; Helmi \& White 2001; Majewski
et al. 2003), Sgr is close to complete disruption, then the youngest
streams should be made up of material which is star dominated, as in
these simulations.  The high mass-to-light ratios suggested by the
observations are most likely due to the fact that the system is not in
virial equilibrium, and its large internal velocity dispersion is
inflated due to the effect of tides, which are strongest close to
pericentre.

To measure the shape of the dark halo of the Galaxy reliably
dynamically old streams are required, and this implies using stars
that trace old stellar populations, such as RR Lyrae stars, the BHBs
and the red giants, as well as main sequence turn-off stars. Such
older streams will typically have very low surface brightness, and
hence are more difficult to detect, particularly if the dark halo of
our Galaxy is not spherical. 

While this paper was being written, Newberg et al. (2003) found an
excess of BHB stars at a distance of 80 kpc from the Sun, in a
direction which lies 10 kpc away from the plane defined by the debris
of Sgr. This prompted them to suggest that it could be material lost
by the dwarf in earlier passages. Notably, Fig.~\ref{fig:dist} of this
Letter shows, for our simulations of the most oblate and most prolate
halos, the presence of debris at a similar location in the sky
($110\ndeg \le \alpha \le 130\ndeg$ and  $\delta \sim 30\ndeg$) and at a
similarly large distance. This debris is old (lost in the first 3
passages) yet it is not far from the great circle defined by the young
debris streams. Thus, even though the Newberg et al. (2003) result
seems at first sight to be very compelling, more careful modeling and
a better estimate of the width of the stream, as well as velocity data
on the BHB star candidates are needed to really constrain the shape of
the dark halo of the Galaxy.  Interestingly the presence of such
ancient streams necessarily imply that Sgr has already been orbiting
the Galaxy for about 10 Gyr.

No other prominent stellar streams have been found thus far in the
Galactic halo. Although the volume probed by the surveys to date has
been somewhat limited, the situation is changing rapidly. The lack of
bright and spatially coherent stellar streams in the Galactic halo
could be suggesting that this is in fact rather flattened.  On the
other hand, it could be indicative of a low satellite accretion rate
or a hint that the majority of the satellites accreted are dark. It is
important to disentangle these issues; they provide direct constraints
both on the nature of dark-matter as well as on the process of galaxy
formation.

\section*{Acknowledgments}
I would like to thank Heather Morrison and Julio Navarro for valuable
discussions that prompted the writing of this paper, and Mariano
M\'endez and Rien van de Weygaert for comments on an earlier version
of this manuscript. The anonymous referee is gratefully acknowledged
for comments which helped improve the paper. This research was funded
through a NOVA fellowship.

\label{lastpage}
\end{document}